# On photoionization of the subvalent subshells of noble gas endohedral atoms

M. Ya. Amusia[1, 2], A. S. Baltenkov[3] and L. V. Chernysheva[2]

[1]Racah Institute of Physics, the Hebrew University, Jerusalem 91904, Israel
[2]Ioffe Physical-Technical Institute, St.-Petersburg 194021, Russia
[3]Arifov Institute of Electronics, Tashkent, 100125, Uzbekistan

**Abstract**

We demonstrate strong interference patterns in the photoionization cross-section of the subvalent subshells of noble gas (NG) endohedral atoms NG@F. This interference is a result of common action of three factors: the effect of neighboring atomic subshells, reflection of photoelectron waves by the fullerene F shell and resonance modification of the incoming photon beam by the complex effect under the action of the F electrons.

We have considered the outer $ns$-subshells for Ne, Ar, Kr and Xe noble gas atoms. The polarization of the fullerene shell is expressed via the F total photoabsorption cross section. The photoelectron reflection from the static F potential is taken into account in the frame of the so-called bubble potential that is a spherical $\delta$-type potential.

It is assumed in the derivations that NG is centrally located in the fullerene. It is assumed also, in accordance with the available experimental data, that the fullerene radius is much bigger than the atomic radius and the thickness of the fullerene shell. These assumptions permit, as demonstrated recently, the NG@F photoionization cross section to be presented as a product of the NG subvalent cross section and two calculated factors that account for polarization of the F electron shell and reflection of photoelectrons by the fullerene static potential.

## 1. Introduction

In this paper we will consider the photoionization of subvalent $ns$-subshells of noble gas (NG) endohedral atoms, formed by a fullerene F, inside which a noble gas atom is embedded, NG@F. We will present data on all the noble gases except He. In concrete calculations, as a fullerene F we will consider $C_{60}$.

The pronounced action of the multi-electron neighboring shell upon a few-electron one was considered for the first time thirty-five years ago. As the first example, the influence of $3p^6$ electrons upon the $3s^2$ in Ar has been presented [1]. A more complicated case with three interacting subshells was considered in [2]. It was demonstrated that the $5p^6$ and $4d^{10}$ subshells act upon $5s^2$ in Xe very strong, completely modifying the $5s^2$ photoionization cross section. All corresponding calculations were performed in the frame of the so-called Random Phase Approximation with Exchange (RPAE). The first experimental confirmations of these predictions were obtained soon [3]. Since then the investigations of the effects of intershell interaction in atoms have become a permanent subject of research (see, e.g. [4, 5]).

The physical nature of these intershell effects in photoionization is as follows. A many-electron atomic subshell is polarized by an electromagnetic wave and a dipole moment is induced in it. Under the action of this dipole moment a neighboring atomic subshell is ionized. RPAE is extremely convenient to describe this effect. So, the ionization of a given electron can proceed via several pathways: directly, after photon absorption by the ionizing electron, and indirectly, in two or even several steps, via virtual excitation of other subshells. Since the electronic subshells in an atom are not separated spatially well enough, the amplitude of these two- or multi-step photo-processes cannot be expressed accurately enough via the dipole polarizability of the many-electron subshells.

In this sense the situation for the endohedral atoms NG@F is quite different. The radius of the fullerene shell significantly exceeds that of an encapsulated atom. This makes it possible for photoionization of the NG atom, in the first approximation, to consider the electronic sub-systems of the fullerene shell and atom as practically independent of each other. For this reason, the amplitude of atom photoionization going through virtual excitation of F shell electrons can be expressed directly via the dynamic polarizability of the fullerene shell $\alpha_{C_{60}}^d(\omega)$. In those cases when the frequency of electromagnetic radiation is close to frequencies of plasma oscillations of the collectivized electrons of the fullerene, the role of this two-step process becomes decisively important, as the role of $4d^{10}$ upon $5s^2$ in isolated Xe.

Along with F shell polarization, one has to take into account also the reflection and refraction of the photoelectron wave, which goes from $ns^2$ subvalent shell, by the static potential of the fullerene. This reflection leads to formation of oscillating pattern of the cross section (see e.g. [6, 7]).

As we will see below, the $ns$ subshell photoionization in the endohedral system NG@F is a remarkable concrete example illustrating the role of the intershell interactions in the fullerene-like molecules, qualitatively similar but even much stronger than in the isolated atoms.

It has been demonstrated recently that the photoionization cross section of the Xe $5s$ subshell is strongly modified due to reflection of the photoelectron wave by the fullerene shell [8]. A simple method was developed to take into account this process. The potential of the $C_{60}$ shell was presented by of a zero-thickness $\delta$-type bubble potential. In this approach the inner degrees of freedom of the $C_{60}$ shell, namely its ability to be polarized, was neglected.

Some time ago we investigated also the role of $C_{60}$ electron shell polarization upon the cross section of $5s^2$ electrons in Xe@$C_{60}$ [9]. We expressed there the effect of the $C_{60}$ shell via the fullerene dipole polarizability and the latter was calculated using considerably simplified expression for the experimental photoabsorption cross section of the $C_{60}$. It appeared, however, that its shape essentially affects the polarizability [10]. That is why here we will not relay on the numerical results in [9] and recalculate them.

Recently, a great deal of attention has been and still is concentrated on photoionization of endohedral atoms. It was demonstrated in a number of papers [11-19] that the $C_{60}$ shell adds prominent resonance structure in the photoionization cross section of endohedral atoms. Although the experimental investigation of A@$C_{60}$ photoionization seems to be very difficult at this moment, these objects will be inevitably intensively studied in the future[1]. This justifies the current efforts of the theorists in predicting rather non-trivial effects waiting for verification.

We will show in this paper that the dynamic polarization of $C_{60}$ drastically modifies the subvalent shell photoionization cross section at any frequency of the incoming radiation $\omega$. The photoionization cross section of subvalent shells of NG endohedral atoms drastically differs from respective data for isolated atoms.

It is of interest to see the alteration of the photoionization cross section if instead of $C_{60}$ other fullerenes, like $C_{70}$, $C_{76}$, $C_{82}$ or $C_{87}$, are considered. However, to study the endohedrals NG@F with F = $C_{70}$, $C_{76}$, $C_{82}$ or $C_{87}$ one needs to know the shape of these objects, their photoionization cross sections and the location of the NG atoms inside the fullerene cage. The answers to these questions are absent at this moment.

---

[1] As a first example of such a research, let us mention the tentative data on measurements of photoionization cross-section of Ce@$C_{82}$ [20].

**2. Essential formulae**

The photoionization characteristics of very many complex atoms were initially calculated only within the framework of RPAE that takes into account along with the direct ionization amplitude of the considered electrons $d_s$ (in our case, these are the subvalent $ns$-electrons) the dipole polarization of other electron shells. The polarized shells ionize the $s$-electron due to inter-shell interaction. The approach developed for isolated atoms can be equally well applied to systems with other electron shells like endohedral atoms.

Symbolically, the total amplitude of some $s$-electron ionization $D_s$ can be presented as a sum of two terms [21]

$$\hat{D}_S = \hat{d}_S + \hat{D}_O \hat{\chi} U_{OS}, \tag{1}$$

where $\hat{D}_O$ is the ionization amplitude of any electrons other than "$s$"-ones, $\hat{\chi} = 1/(\omega - \hat{H}_{ev}) - 1/(\omega + \hat{H}_{ev})$ is the propagator of other electron excitation, i.e. electron-vacancy pair creation, $\hat{H}_{ev}$ is the pair Hartree-Fock Hamiltonian and $U_{OS} \equiv V_{OS,dir} - V_{OS,exch}$, with $V_{OS,dir}$ and $V_{OS,exch}$ being the operator of direct and exchange pure Coulomb interaction between "$o$" and "$s$" electrons.

We concentrate on almost spherically symmetric systems. The formula (1) is simplified considerably if the "$o$"-electrons are either at much smaller distances or much larger ones from the center of the system than the "$s$"-electrons. In both cases the Coulomb interaction is considerably simplified, becoming either

$$U_{OS} \approx \frac{1}{r_S} - \frac{\mathbf{r}_O \cdot \mathbf{r}_S}{r_S^3}, \text{ (for } r_S \gg r_O\text{)}, \tag{2a}$$

or

$$U_{OS} \approx \frac{1}{r_O} - \frac{\mathbf{r}_O \cdot \mathbf{r}_S}{r_O^3}, \text{ (for } r_O \gg r_S\text{)}. \tag{2b}$$

Here $\mathbf{r}_s$ and $\mathbf{r}_o$ are the "$s$"- and "$o$"-electron shells radii, respectively.

The equation (1) can be easily generalized in the spirit of the Landau Fermi-liquid theory by incorporating into $d_s$ all but simple electron-vacancy excitations, for example, "two electron – two vacancy" excitations of the $s$-shell [22].

The effect of the "$o$"-shell is presented particularly simple when it is an inner one located well inside the intermediate and outer atomic subshell. Then rightfully neglecting the exchange "$o$-$s$"-interaction and representing $U_{OS}$ as (2a), one reduces (1) to an algebraic equation instead of operator one where $\hat{D}_O \hat{\chi} U_{OS}$ is substituted by the following expression

$$[2 \sum_{evexit,O} \omega_{ev} D_{ev}(\omega)(\omega^2 - \omega_{ev}^2)^{-1} d_{ev}]/r_S^3 \equiv -\alpha_O^d(\omega)/r_S^3. \tag{3}$$

Here it is taken into account that the first term of expansion (2a) makes no contribution to the Coulomb matrix element because of orthogonality of the atomic wave functions. The summation over $evexit, o$ includes all electron-vacancy excitation of the considered shell. In

(3) we use an alternative definition of the dipole polarizability $\alpha_o(\omega)$ of the "$o$"-shell. Usually, it is defined as

$$\alpha_O^d(\omega) \equiv -[2 \sum_{evexit,o} \omega_{ev} \mid D_{ev}(\omega_{ev}) \mid^2 (\omega^2 - \omega_{ev}^2)^{-1}], \qquad (4)$$

but it can be demonstrated that this definition and that in (4) are identical (see [22] and references therein).

Thus, in the case of an inner shell "$o$" one has instead of Eq. (1) the following formula [9]:

$$D_S(\omega) \cong d_S \left(1 - \frac{\alpha_O^d(\omega)}{r_S^3}\right), \qquad (5)$$

Usually, of interest is $D_s(\omega)$ for photon energy $\omega^2$ of an order of $I_s$ being the $s$-electron ionization potential. The outer shell photoionization cross section has its highest values at $\omega \sim I_s$. Since $I_s \ll I_o$, one can substitute $\omega$ in $\alpha_o^d(\omega)$ of (5) by zero, having instead of dynamic $\alpha_o^d(\omega)$ the static dipole polarizability $\alpha_o^d$.

If one considers as "$o$" the outer electrons with $r_o \gg r_s$ the expression Eq. (1) is again considerably simplified. As above, the exchange can be again neglected. Usually, the "$o$"-shell is a layer of electrons the thickness of which is much smaller than its radius $r_o$. In this case one obtains from consideration performed in [9] an expression similar to Eq. (5) where $r_s^3$ is substituted by a mean value $\bar{r}_o^3$:

$$D_S(\omega) \cong d_S \left(1 - \frac{\alpha_O^d(\omega)}{\bar{r}_O^3}\right). \qquad (6)$$

For $\omega$ above the $s$-subshell ionization threshold, since for $\omega > I_s \gg I_o$ the polarizability $\alpha_o^d(\omega)$ from Eq. (4) can be simplified by neglecting $\omega_{ev}$ as compared to $\omega$, and thus substituted by its dynamic high $\omega$ limit, $\alpha_o^d(\omega) = -N_O/\omega^2$. This correction for an isolated atom is small since $\omega^2 \sim I_S^2 \sim N_S/r_S^3$, that means substituting $\alpha_o^d(\omega)/\bar{r}_O^3$ by $N_O r_S^3 / N_S r_O^3 \ll 1$.

For isolated atoms Eq. (1) has to be solved numerically, without additional simplifications connected to the smallness of the ratios $r_s/r_o$ or $r_o/r_s$. Then the effect of any additional electrons, e. g. those belonging to the $C_{60}$ shell in endohedral atoms can be treated as something on top of the atomic multi-electron effects. An essential simplification comes from the fact, that the $C_{60}$ radius $R_c$ is big enough, $R_C \gg r_S$. It is also essential that the electrons in $C_{60}$ are located within a layer the thickness of which $\Delta R_c$ is considerably smaller than $R_c$. In this case, one comes to an expression similar to Eq. (5), except when $d_s$ is substituted by $D_S^{(A)}(\omega)$, i.e. by the amplitude of $s$-electron photoionization with all essential atomic correlations taken into account:

---

[2] The atomic system of units: $e = m = \hbar = 1$ is used throughout this paper.

$$D_S(\omega) \cong D_S^{(A)}(\omega)\left(1 - \frac{\alpha_O^d(\omega)}{R_c^3}\right). \tag{7}$$

Using the relation between the imaginary part of the polarizability and the total photoabsorption cross-section, one can derive the polarizability of the C$_{60}$ shell as any other object. This relation looks as follows: $\text{Im}\,\alpha_{C_{60}}^d(\omega) = c\sigma_{C_{60}}(\omega)/4\pi\omega$. Although experiments (see [23] and references therein) provide no direct absolute values of $\sigma_{C_{60}}(\omega)$, it can be reliably estimated theoretically using different normalization procedures on the basis of the sum rule: $(c/2\pi^2)\int_{I_o}^{\infty}\sigma_{C_{60}}(\omega)d\omega = N$, where $N$ is the number of collectivized electrons. This approach was used for polarizability of C$_{60}$ in [10], where it was considered that $N = 240$, i.e. 4 collectivized electrons per each C atom in C$_{60}$. The real part of polarizability is connected with the photoabsorption cross section by the dispersion relation

$$\text{Re}\,\alpha_{C_{60}}^d(\omega) = \frac{c}{2\pi^2}\int_{I_{60}}^{\infty}\frac{\sigma_{C_{60}}(\omega')d\omega'}{\omega'^2 - \omega^2}, \tag{8}$$

where $I_{60}$ is the C$_{60}$ ionization potential, $c$ is the light of speed. We omitted in this expression the terms connected with discrete excitations of fullerene since the oscillator strengths corresponding to the electron discrete transitions are small.

In order to take into account the processes of reflection and refraction of the photoelectron wave by the fullerene shell, we use a $\delta$-bubble model that represents the static C$_{60}$ potential as $U(r) = -V_0\delta(r - R_c)$ with $V_0$ chosen in such a way as to reproduce the binding energy of $C_{60}^-$ negative ion. In the frame of this model potential, its influence upon the photoionization amplitude is presented by a factor $F_A(k)$ that takes into account reflection of the $p$-photoelectron by the C$_{60}$ shell. The details of calculation of $F_A(k)$ can be found in [8]. In short, this function is expressed via the regular and irregular photoelectron wave functions:

$$F_l(k) = \cos\Lambda_l(k)\left[1 - \tan\Lambda_l(k)\frac{v_{kl}(R_c)}{u_{kl}(R_c)}\right], \tag{9}$$

where $\Lambda_l(k)$ are the additional phase shifts due to the static fullerene shell potential. They are expressed by the following formula:

$$\tan\Lambda_l(k) = \frac{u_{kl}^2(R_c)}{u_{kl}(R_c)v_{kl}(R_c) + k/2V_0}. \tag{10}$$

In these formulas $u_{kl}(r)$ and $v_{kl}(r)$ are the regular and irregular solutions of the atomic Hartree-Fock (H-F) equations for a photoelectron with momentum $k = \sqrt{2\varepsilon}$, where $\varepsilon$ is the photoelectron energy connected with the photon energy $\omega$ by the relation $\varepsilon = \omega - I_A$ with $I$ being the atom A ionization potential.

Entirely, the following relation gives the amplitude $D_S(\omega)$ for photoionization of the *ns*-electrons of endohedral atom:

$$D_S(\omega) = F_A(k) D_S^{(A)}(\omega) \left(1 - \frac{\alpha_{C_{60}}^d(\omega)}{R_C^3}\right) \equiv F_A(k) D_S^{(A)}(\omega) G_C^d(\omega), \quad (11)$$

where the factor $G_C^d(\omega)$ is a complex function, $G_C^d(\omega) \equiv \tilde{G}_C^d(\omega) \exp[i\eta^d(\omega)]$, $\tilde{G}_C^d(\omega)$ being the modulus of $G_C^d(\omega)$.

Using this amplitude, one has for the cross section

$$\sigma_S(\omega) = \sigma_S^{(A)}(\omega) F_A^2(\omega) \left|1 - \frac{\alpha_{C_{60}}^d(\omega)}{R_C^3}\right|^2 \equiv \sigma_S^{(A)}(\omega) F_A^2(\omega) S_C^d(\omega), \quad (12)$$

where $S_C^d(\omega) = [\tilde{G}_C^d(\omega)]^2$ can be called radiation enhancement parameter.

If one has the dipole and quadrupole photoionization amplitudes, one can obtain the photoionization differential cross section. It is given by the following formula valid for linearly polarized radiation (see reference in e.g. [21]):

$$\frac{d\sigma_{ns}(\omega)}{d\Omega} = \frac{\sigma_{ns}(\omega)}{4\pi}[1 + 2P_2(\cos\Theta) + \gamma_{ns}^C(\omega)\cos^2\Theta \sin\Theta \cos\Phi]. \quad (13)$$

Here $\Theta$ is the polar angle between the vectors of photoelectron velocity $\mathbf{v}$ and photon polarization $\mathbf{e}$, while $\Phi$ is the azimuthal angle determined by the projection of $\mathbf{v}$ in the plane orthogonal to $\mathbf{e}$ that includes the vector of photon's velocity, $P_2(\cos\Theta)$ is the Legendre polynomial.

The non-dipole parameter $\gamma_{ns}^C(\omega)$ in one-electron Hartree-Fock approximation has the form:

$$\gamma_{ns}^C(\omega) = 6\frac{\omega}{c}\frac{q_2}{d_1}\cos(\delta_2 - \delta_1). \quad 14)$$

Here $\delta_l \equiv \delta_l(k)$ are the photoelectron scattering phases; the following relation gives the matrix elements $d_1$ in the so-called *r*-form

$$d_1 \equiv \int_0^\infty P_{ns}(r) r P_{\varepsilon p}(r) dr, \quad (15)$$

where $P_{ns}(r)$ and $P_{\varepsilon p}(r)$ are the radial H-F [23, 24] one-electron wave functions of the *ns* discrete level and *εp* continuous spectrum, respectively. The following relation gives the quadrupole matrix elements

$$q_2 \equiv \frac{1}{2}\int_0^\infty P_{ns}(r) r^2 P_{\varepsilon d}(r) dr. \quad (16)$$

In order to take into account the Random Phase Approximation with Exchange (RPAE) multi-electron correlations [21], one has to perform the following substitutions in the expressions for $\gamma_{ns}^{C}(\omega)$ [21]:

$$d_1 q_2 \cos(\delta_2 - \delta_1) \to \tilde{D}_1 \tilde{Q}_2 \cos(\delta_2 + \Delta_2 - \delta_1 - \Delta_1),$$
$$d_1^2 \to \tilde{D}_1^2 \qquad (17)$$

Here the following notations are used for the matrix elements with account of multi-electron correlations, dipole and quadrupole, respectively:

$$D_1(\omega) \equiv \tilde{D}_1(\omega) \exp[i\Delta_1(\omega)]; \ Q_2(\omega) \equiv \tilde{Q}_2(\omega) \exp[i\Delta_2(\omega)], \qquad (18)$$

where $\tilde{D}_1(\omega)$, $\tilde{Q}_2(\omega)$ are the absolute values of the amplitudes for dipole and quadrupole transitions; $\Delta_1(\omega)$ and $\Delta_2(\omega)$ the arguments of these amplitudes.

The following expressions are the RPAE equations for the dipole matrix elements

$$\langle v_2|D(\omega)|v_1\rangle = \langle v_2|d|v_1\rangle + \sum_{v_3,v_4} \frac{\langle v_3|D(\omega)|v_4\rangle(n_{v_4} - n_{v_3})\langle v_4 v_2|U|v_3 v_1\rangle}{\varepsilon_{v_4} - \varepsilon_{v_3} + \omega + i\eta(1 - 2n_{v_3})}, \qquad (19)$$

where

$$\langle v_1 v_2|\hat{U}|v_1' v_2'\rangle \equiv \langle v_1 v_2|\hat{V}|v_1' v_2'\rangle - \langle v_1 v_2|\hat{V}|v_2' v_1'\rangle. \qquad (20)$$

Here $\hat{V} \equiv 1/|\vec{r} - \vec{r}'|$ and $v_i$ is the total set of quantum numbers that characterize a H-F one-electron state on discrete (continuum) levels. This set includes the principal quantum number (energy), angular momentum, its projection and the projection of the electron spin. The function $n_{v_i}$ (the so-called step-function) is equal to 1 for an occupied state and 0 for a vacant one; $\eta \to +0$.

The dipole matrix elements $D_1$ are obtained by solving the radial part of the RPAE equation (12). As to the quadrupole matrix elements $Q$, they are obtained by solving the radial part of the RPAE equation similar to (19)

$$\langle v_2|Q(\omega)|v_1\rangle = \langle v_2|\hat{q}|v_1\rangle + \sum_{v_3,v_4} \frac{\langle v_3|Q(\omega)|v_4\rangle(n_{v_4} - n_{v_3})\langle v_4 v_2|U|v_3 v_1\rangle}{\varepsilon_{v_4} - \varepsilon_{v_3} + \omega + i\eta(1 - 2n_{v_3})}. \qquad (21)$$

Here in the $r$-form one has $\hat{q} = r^2 P_2(\cos\theta)$. Equations (19, 21) are solved numerically using the procedures discussed at length in [24].

Inclusion of polarization of the fullerene shell by incoming light and reflection of the photoelectrons leads to (11) for the dipole and to the following expressions for quadrupole $Q_{nl,nl'}^{AC}(\omega)$ photoionization amplitude:

$$Q_{nl,nl'}^{AC}(\omega) \cong F_{l''}(k)Q_{nl,nl'}^{A}(\omega)\left(1 - \frac{\alpha_{C_{60}}^{q}(\omega)}{R_C^5}\right) \equiv F_{l''}(k)Q_{nl,nl'}^{A}(\omega)G_{C_{60}}^{q}(\omega), \quad (22)$$

where $\alpha_{C_{60}}^{q}(\omega)$ is the quadrupole dynamical polarizability of C$_{60}$. The factor $G_{C_{60}}^{q}(\omega)$ is a complex function and as such can be presented as

$$G_{C_{60}}^{q}(\omega) = G_{C_{60}}^{q}(\omega)\exp[i\Lambda^{d,q}(\omega)]. \quad (23)$$

With the amplitudes (11, 22), the non-dipole parameter $\gamma_{ns}^{C}(\omega)$ becomes equal to the expression that follows from (14, 18, 11, 22):

$$\gamma_{ns}^{C}(\omega) = 6\frac{\omega}{c}\frac{\tilde{Q}_2(\omega)\tilde{G}_C^q(\omega)F_2(k)}{\tilde{D}_1(\omega)\tilde{G}_C^d(\omega)F_1(k)}\cos(\tilde{\delta}_2 - \tilde{\delta}_1). \quad (24)$$

where $\tilde{\delta}_{2,1} = \delta_{2,1} + \Delta_{2,1} + \eta^{q,d}$ (see (18)).

Preliminary investigations demonstrated that $G_{C_{60}}^{q}(\omega)$ is close to unity. That means that the role of quadrupole polarization can be neglected. This is why in the below-presented results we assume that $G_{C_{60}}^{q}(\omega) = 1$ and $\eta^q = 0$. We plan to pay special attention to the role of quadrupole excitations in general and quadrupole continuous spectrum resonances in particular in the future. However without experimental data this task is far from being trivial.

**3. Results of calculations**

The C$_{60}$ parameters in the present calculations were chosen the same as in the previous papers, e.g. in [9]: $R = 6.639$ and $V_0 = 0.443$.

In Fig. 1 we present the radiation enhancement parameter $S(\omega)$ (see (12)), its amplitude's absolute value $\tilde{G}^d(\omega) \equiv |G(\omega)|$ and phase $\eta^d \equiv \arg G(\omega)$. Although the curves are rather complex, the $ns$ thresholds for Ne, Kr and Xe are located at such energies that only smoothly decreasing part of $S(\omega)$ can affect the photoionization cross-section and non-dipole angular anisotropy parameter. The situation for Xe is different, so the enhancement factor acts a little bit trickier there.

In Fig. 2-5 we depict data for the photoionization cross-section as well as non-dipole angular anisotropy parameter of subvalent $ns^2$-subshell in NG@C$_{60}$, where NG = Ne, Ar, Kr, Xe.

In all considered cases we see prominent influence of the fullerenes shell upon the photoionization of the "caged" atoms – Ne, Ar, Kr and Xe. Rather impressive are the oscillations due to reflection of the photoelectron by the fullerenes shell. As to the influence of the radiation enhancement parameter, it is not too big in Ne and steadily increasing toward Xe. For Ne, with its relatively high ionization potential of 53 eV the effect of G($\omega$) and $S(\omega)$ is noticeable only at threshold. Already in Ar@C$_{60}$ the first maximum in the cross-section became bigger due to $S(\omega)$ by a factor of more than three. Prominently increases the second maximum. Noticeable is the decrease in the non-dipole parameter (see Fig. 3). As is seen from Fig.4 and location of the 4s threshold in Kr, the effects of $S(\omega)$ and G($\omega$) are even stronger here than in Ar. According to Fig. 5 and as a consequence of location of the 5s threshold 26 eV almost at the maximum in $S(\omega)$ and $G(\omega)$, the effects of radiation enhancement factor are here bigger than in all other noble gases. Note, however, that at the very threshold in Xe, the corrections due to reflection and radiative enhancement almost compensate each other.

Although the role of $S(\omega)$ and $G(\omega)$ for subvalent shells are not as big as for the outer *np* [25], it is still noticeable and of great interest.

## 4. General discussion

We considered above as a fullerene only $C_{60}$. As it was mentioned in the introduction, it would be of interest to see the alteration of the photoionization cross-section if instead of $C_{60}$ other fullerenes, like $C_{70}$, $C_{76}$, $C_{82}$ or $C_{87}$ are considered. We do not know the shape and photoionization cross-sections of $C_{70}$, $C_{76}$, $C_{82}$ or $C_{87}$ and the position of the NG atoms inside the fullerenes. However, to have the feeling of the fullerenes shell effect upon photoionization of NG, we can properly use the results for $C_{60}$ by scaling them to other radius, collectivized electrons number etc. Since the effects of radiative enhancement and oscillations due to reflection are sensitive to the radius, potential of the fullerene and number of electrons in it, one can expect

It is essential to have in mind that while being caged, the atoms inside can be ionized. The electrons go to the fullerenes shell that became instead of a neutral, a negatively charged surface. This requires a modification in the accounting for the reflection of the photoelectron by the fullerenes shell. The latter cannot be considered as neutral and described by a zero-thickness potential, but instead by a combination of it with a Coulomb long range potential. Such a modification, although straightforward, makes the calculation procedure that leads to a rather simple $F_l(k)$ factor used in this paper, considerably more complex.

**Acknowledgements**

MYaA is grateful for financial support to the Israeli Science Foundation, Grant 174/03 and the Hebrew University Intramural Funds. ASB expresses his gratitude for financial support by Uzbekistan National Foundation, Grant ФА-Ф4- Ф095.

**References**

1. M. Ya. Amusia, V. K. Ivanov, N. A. Cherepkov and L. V. Chernysheva, Phys. Lett. A **40**, 5, 361 (1972).
2. M. Ya. Amusia, N. A. Cherepkov, Case Studies in Atomic Physics, North-Holland Publishing Company, **5**, 2, p. 47-179, 1975.
3. J. A. R. Samson and J. L. Gardner, Phys. Rev. Lett. **33**, 671 (1974).
4. H. Kjeldsen, P. Andersen, F. Folkmann, H. Knudsen, B. Kristensen, J. B. West, T. Andersen, Phys. Rev. A **62**, 020702(R) (2000).
5. K. Koizumi *et al*, Phys. Scr. **T73**, 131 (1977).
6. A. S. Baltenkov, Phys. Lett. A **254**, 203 (1999); J. Phys. B **32**, 2475 (1999).
7. M. Ya. Amusia, A. S. Baltenkov, V. K. Dolmatov, S. T. Manson, and A. Z. Msezane, Phys. Rev. A **70**, 023201 (2004).
8. M. Ya. Amusia, A. S. Baltenkov, and U. Becker, Phys. Rev. A **62**, 012701 (2000).
9. M. Ya. Amusia and A. S. Baltenkov, Phys. Rev. A **73**, 062723, 2006.
10. M. Ya. Amusia and A. S. Baltenkov, Phys. Lett. A **360**, 294-298, 2006.
11. M. J. Pushka and R. M. Niemenen. Phys. Rev. B **47**, 1181 (1993).
12. G. Wendin and B. Wastberg. Phys. Rev. B **48**, 14764 (1993).

13. L. S. Wang, J. M. Alford, Y. Chai, M. Diener, and R. E. Smalley. Z. Phys. D. **26**, S297 (1993).
14. P. Decleva, G. De Alti, M. Stener. J. Phys. B **32**, 4523 (1999).
15. J.-P. Connerade, V. K. Dolmatov, and S. T. Manson. J. Phys. B **33**, 2279 (2000).
16. J. P. Connerade, V. K. Dolmatov, and S. T. Manson J. Phys. B **33**, L275 (2000).
17. H. Shinohara. Rep. Prog. Phys. **63**, 843 (2000).
18. M. Stener, G. Fronzoni, D. Toffoli, P. Colavita, S. Furlan, and P. Decleva. J. Phys. B **35**, 1421 (2002).
19. J. Kou, T. Mori, M. Ono, Y. Haruyama, Y. Kubozono, and K. Mitsuke. Chem. Phys. Lett. **374**, 1 (2003).
20. R. Phaneuf, private communication (2007).
21. M. Ya. Amusia, Radiation Physics and Chemistry, **70**, 237-251 (2004).
22. M. Ya. Amusia, in: *Photoionization in VUV and Soft X-Ray Energy Region*, ed. U. Becker and D. Shirley, New York - London, Plenum Press, p. 1-46 (1996).
23. M. Ya. Amusia, *Atomic Photoeffect* (Plenum Press, New York – London, 1990).
24. M. Ya. Amusia and L.V. Chernysheva, *Computation of Atomic Processes* ("Adam Hilger" Institute of Physics Publishing, Bristol – Philadelphia, 1997).
25. M. Ya. Amusia, A. S. Baltenkov, and L.V. Chernysheva, Phys. Rev. Lett., submitted, 2007 http://arxiv.org/abs/0707.4404; Phys. Rev. A, submitted, 2007 http://arxiv.org/abs/0710.3910

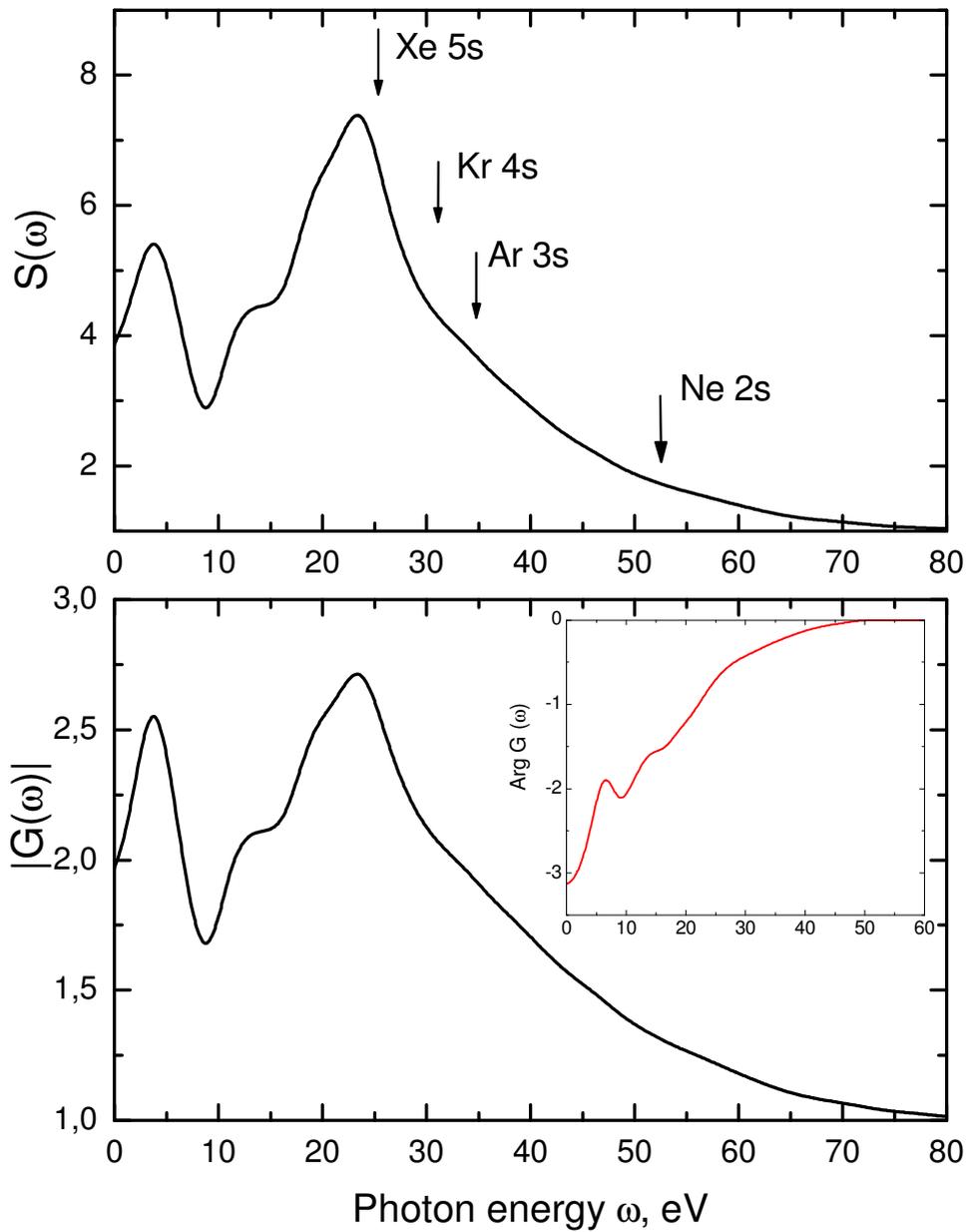

Fig.1. Radiation enhancement parameter $S(\omega)$, its amplitude's absolute value $\tilde{G}^d(\omega) \equiv |G(\omega)|$ and phase $\eta^d \equiv \arg G(\omega)$. Arrows denote the thresholds positions of corresponding subvalent *ns* subshells.

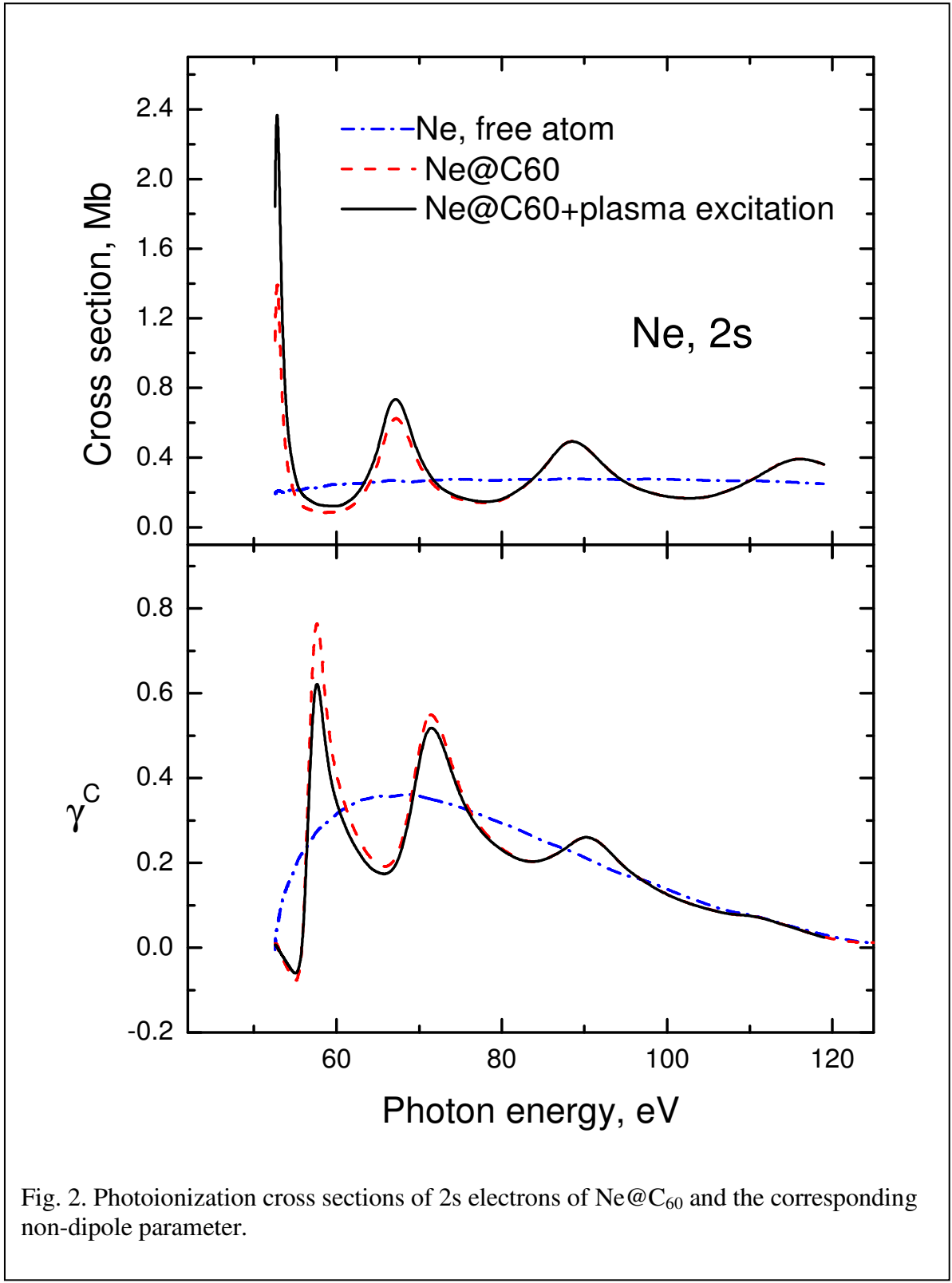

Fig. 2. Photoionization cross sections of 2s electrons of Ne@C$_{60}$ and the corresponding non-dipole parameter.

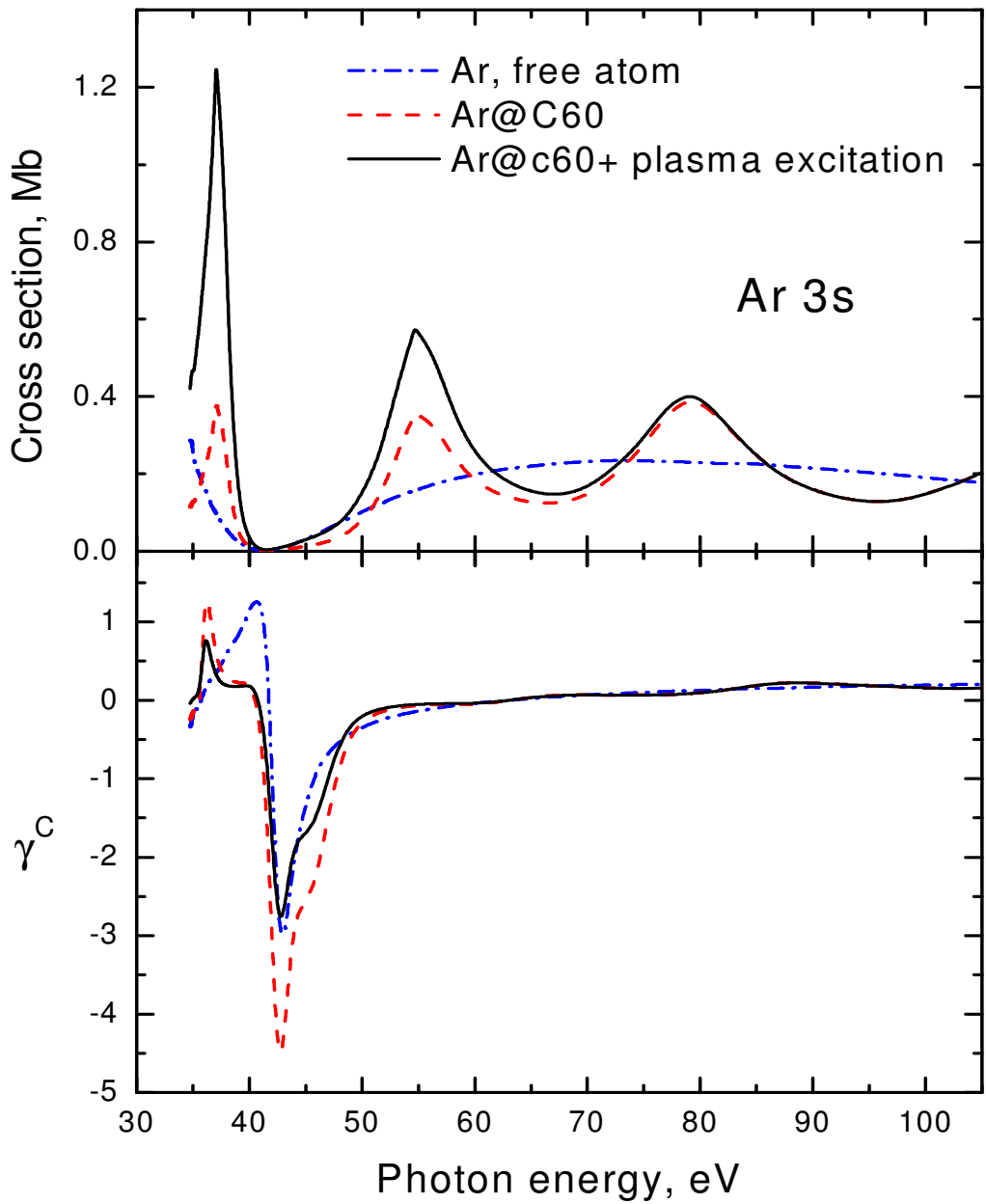

Fig. 3. Photoionization cross-sections of 3s electrons of Ar@$C_{60}$ and the corresponding non-dipole parameter.

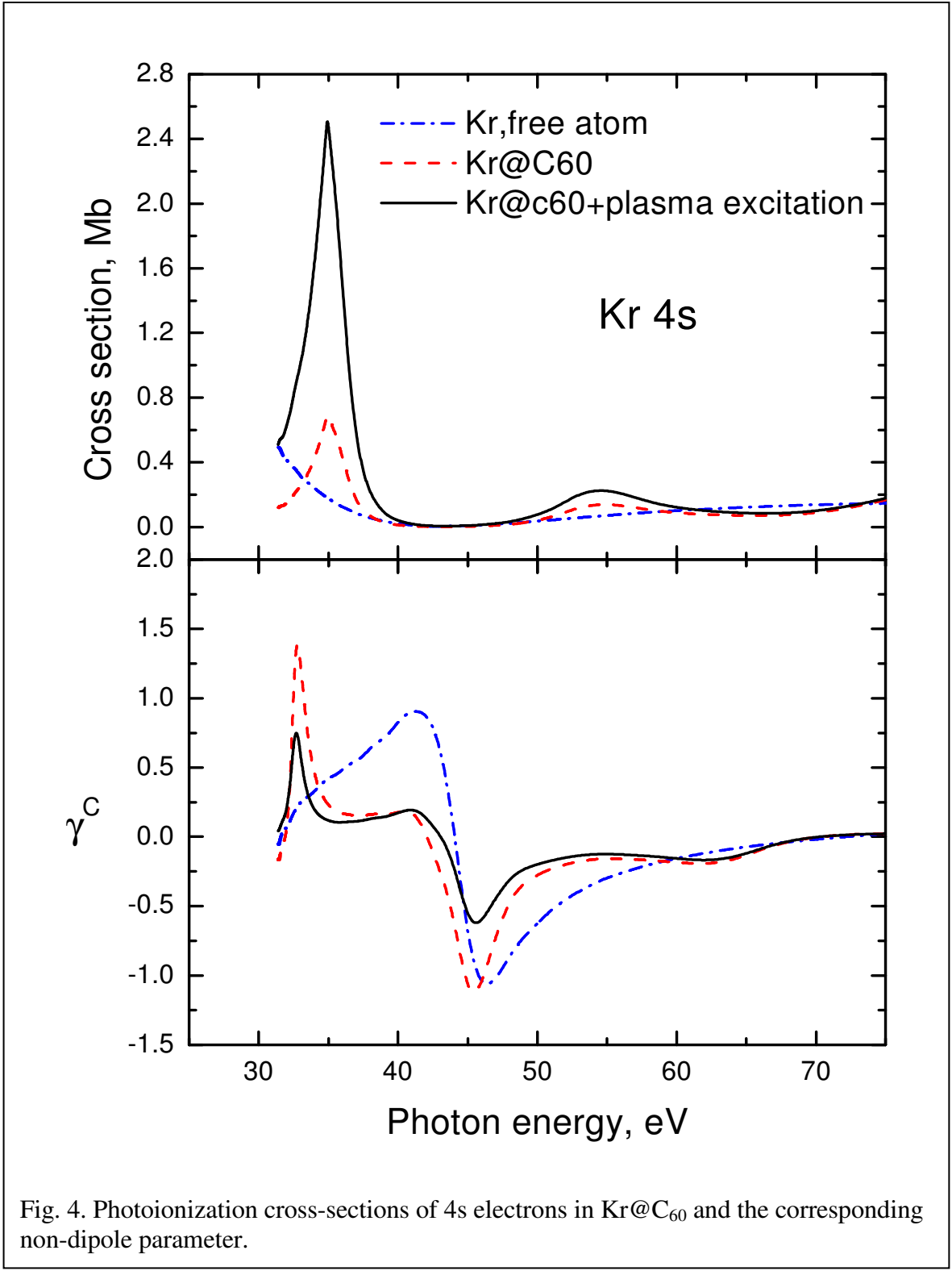

Fig. 4. Photoionization cross-sections of 4s electrons in Kr@C$_{60}$ and the corresponding non-dipole parameter.

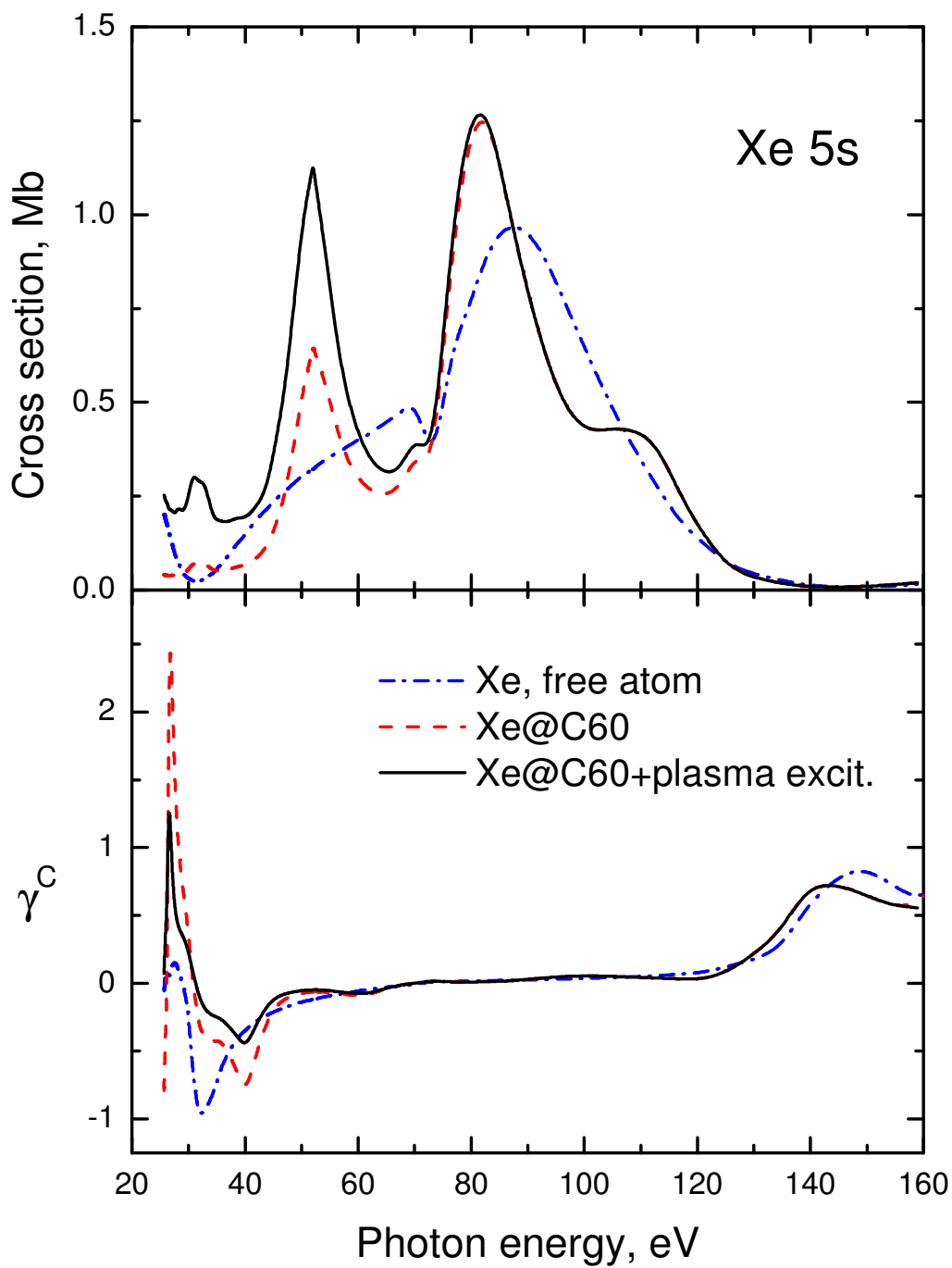

Fig. 5. Photoionization cross-sections of 4s electrons in Xe@$C_{60}$ and the corresponding non-dipole parameter.